\documentclass[]{elsart3p}

\usepackage{dcolumn}
\usepackage{bm}
\usepackage{graphicx}        
\usepackage{cite}            
\usepackage{bm}
\usepackage{amsmath}
\usepackage{amssymb}

\usepackage{amssymb}

\journal{Physica C}

\begin{document}

\begin{frontmatter}



\title{Self-consistent Ginzburg-Landau theory for transport currents in superconductors} 



\author{M.~\"{O}gren, M.~P.~S\o rensen and N.~F.~Pedersen}

\address{Dep. of Mathematics, Technical University of Denmark, 2800 Kongens Lyngby, Denmark}

\begin{abstract}

We elaborate on boundary conditions for Ginzburg-Landau (GL) theory in the case of external currents. 
We implement a  self-consistent theory within the finite element method (FEM) and present numerical results for a two-dimensional rectangular geometry.
We emphasize that our approach can in principle also be used for general geometries in three-dimensional superconductors.

\end{abstract}

\begin{keyword}

Ginzburg-Landau theory \sep Transport  \sep Vortices  \sep Self-consistent boundary conditions  \sep FEM


\end{keyword}

\end{frontmatter}


\section{Introduction}
\label{}

There is an increasing interest of using superconductors for example in the design of products where a reduced size and weight is crucial, such as generators in the nacelle of large off shore wind turbines and electrical motors in azimuth thrusters for ships \cite{Seiler2010}.
Modeling of superconductors for technological applications is a truly multiscale problem.
The remarkable development of computers for scientific calculations 
makes it practically possible to take further steps in the sophistication of quantitative modeling of superconductors with transport currents.
FEM modeling of superconducting tapes in realistic technological devices can today be performed under the assumption of an empirical expressions for the resistivity \cite{VictorIEEE2011}.
A magnetic field is generally expelled from the interior of a superconductor.
However, in type-II superconductors, quantized magnetic flux can penetrate from the surface and form vortices which increase the macroscopic resistivity. 
The Ginzburg-Landau (GL) theory is a celebrated tool for theoretical modelling of superconductors  \cite{Ginzburg1950, Schmid1966, Gorkov1968} such as for example of vortex dynamics in type-II superconductors, see e.g. \cite{KatoPRB1991}. 
The focus in the present paper is to discuss implementations of GL theory for the microscopic modelling of currents through type-II superconductors. 

\section{Theoretical formalism}
\label{}

The time-dependent GL theory is strictly valid for gapless superconductors and for temperatures close to $T_c$, but have been confirmed experimentally for a wide range of temperatures \cite{Tinkham}.
We do not study the temperature dependence here,
moreover we fix the electromagnet gauge to be the (incomplete) {\it Weyl gauge}, i.e. the electric scalar potential is zero $\varphi=0$ througout the paper.
The dimensionless time-dependent Ginzburg-Landau equations to be solved self-consistently are 

\begin{eqnarray}
\label{GLE}
\begin{array}{l}
\partial_t \psi=-\left(\frac{ i}{ \kappa }\nabla+\mathbf{A}\right)^{2}\psi+\psi - \left|\psi\right|^{2}\psi,
\\
\sigma\partial_t \mathbf{A}= \frac{ 1}{ 2i\kappa } \left(\psi^{*}\nabla\psi-\psi\nabla\psi^{*}\right) -\left|\psi\right|^{2}\mathbf{A}   
\\
\ \ \ \ \ \ \ \ \ \ -\nabla\times \left( \nabla\times\mathbf{A} -\mathbf{B}^{(a)} \right), \  \mathbf{r} \in \Omega;
\\
\nabla \times \mathbf{A}|_{  \mathbf{r} \in \delta \Omega} = \mathbf{B}^{(a)} + \frac{1}{4\pi} \int_{\Omega} \frac{ \mathbf{J}\left(\mathbf{r}'\right)\times\left(\mathbf{r}-\mathbf{r}'\right) }{ \left|\mathbf{r}-\mathbf{r}'\right|^{3} } dV'. 
\end{array}
\end{eqnarray}
Here, $\psi$ is the order-parameter for the superconducting Cooper-pair condensate, while $\mathbf{A}$ is the magnetic vector potential.
Moreover, $\mathbf{B}^{(a)} \left(\mathbf{r}\right)$ is the dimensionless applied magnetic field, while the integral operator in the lower equation is the  current-induced magnetic field following \emph{Biot-Savarts law}, where $  \mathbf{J} =  \mathbf{J}^{(s)}  +  \mathbf{J}^{(n)}$ is the sum of  \emph{super-currents} and \emph{normal-currents} \cite{Tinkham}
\begin{equation}
 \mathbf{J}^{(s)}=\frac{i}{2\kappa} \left(  \psi \nabla \psi^*  - \psi^* \nabla \psi     \right) -\left|\psi\right|^{2} \mathbf{A}; 
\  \mathbf{J}^{(n)}=  - \sigma  \partial_t  \mathbf{A}.  \label{Js}
\end{equation}
Note that the last term in the lower equation in (\ref{GLE}) differs from the  \emph{standard} time-dependent GL equations, but is  instrumental in incorporating transport currents as we will detail below.
The parameters in the problem are $\sigma,\:\kappa, \: I$ and $ \mathbf{B}^{(a)}$.
They are related to the (unscaled) physical quantities  ($P$) as follows:
$\sigma_P=\sigma/\left(\mu_{0}D\kappa^{2}\right)$, where $\sigma$
is the conductivity of the normal current, $\mu_{0}$
is the permeability of vacuum, $D$ is a phenomenological diffusion
coefficient \cite{Gorkov1968}; 
$\kappa=\lambda/\xi$ is the \emph{Ginzburg-Landau
parameter} (approximately independent of temperature), it is the ratio between the \emph{London penetration
depth} $\lambda$ for external magnetic fields and the coherence length  $\xi$ of the (Cooper-pair) condensate.
In the present formulation we use $\lambda$ as lengthscale, while time is scaled according to $\xi^2/D$.
The transport current to be modeled is $I_P= \hbar I / \left(\mu_0 q\xi  \lambda  \right) $, where $\hbar$
is Plancks constant divided by $2\pi$ and $q=2e$ is the electric charge of a Cooper pair. 
The magnetic vector potential is scaled according
to $\mathbf{A}_P=\hbar\mathbf{A}/\left(q\xi\right)$, hence the magnetic field is $\mathbf{B}_P=\nabla \times \mathbf{A}_P =\hbar\mathbf{B}/\left(q\xi \lambda \right)$.

We now discuss the qualitative behaviour of currents within the superconductor in general terms. 
The super-currents $\mathbf{J}^{(s)}$ is zero at the points in contact with the normal conductors.
Super-currents carry the main part of the transport current within the superconductor, except in the vicinity of vortices where super-currents encircle the centers of the vortices (see Fig.~\ref{figure1}). 
The normal-currents $\mathbf{J}^{(n)}$ carry the total current close to the normal conductors.
Normal-currents are also present at positions where the Cooper-pair density is penetrated by moving vortices, the normal-currents here are responsible for dissipation inside the superconductor and build up a macroscopic resistance.
Hence from a technological point of view, it is important to optimize geometries, dimensions, material properties and pinning of vortices, in order to obtain desirable current-voltage relations for superconductors.
However, in general there are rather few articles treating GL theory for transport currents, see \cite{MachidaPRL1993, GroppJoCP1996, WinieckiPRB2002, VodolazovPhysicaC2004} for examples.

Crucial for the modeling of transport is how to set up the boundary conditions (BC). 
From a physical point of view, it involves a condition for the super-currents $\mathbf{J}^{(s)} \cdot \mathbf{n}=0$ not to penetrate any boundary,
a specification of the magnetic field $\mathbf{B}$ at the boundary,
and conditions for the normal-currents $\mathbf{J}^{(n)}$ at the boundaries.
 The \emph{vacuum-superconductor BC} for the standard time-dependent GL equations are  \cite{VodolazovPhysicaC2004} 
\begin{equation}
\left(  \frac{i}{\kappa} \nabla + \mathbf{A} \right)\psi\cdot\mathbf{n}=0; \ \nabla\times\mathbf{A}=\mathbf{B}^{(e)}; \ -\sigma \partial_t \mathbf{A} \cdot\mathbf{n} = 0 , 
\label{VSBC}
\end{equation}
while for regions where the current is injected, the  \emph{metal-superconductor BC} are
\begin{equation}
\psi=0; \ \nabla\times\mathbf{A}=\mathbf{B}^{(e)}; \ -\sigma \partial_t \mathbf{A} \cdot\mathbf{n} = \mathbf{J}^{(e)} \cdot\mathbf{n}. 
\label{MSBC}
\end{equation}
Here, $\mathbf{n}$ is the outgoing normal vector to $\partial\Omega$ and $ \mathbf{J}^{(e)} $ is the \emph{external} current density.
The \emph{external} magnetic field is the sum of the \emph{applied} and \emph{current-induced} fields $\mathbf{B}^{(e)}=\mathbf{B}^{(a)}+ \mathbf{B}^{(c)}$ as shown in the lower equation~(\ref{GLE}).
Note that the applied field may result from other superconductors, for example in a generator or electric motor.
In a general three-dimensional geometry the implementation of the presented theory may be challenging and computationally very demanding.
Therefore we here give results for illustrative purposes of a standard geometry in numerical GL simulations, the two-dimensional ($z=0$) rectangle. 
That is a superconductor of  length $L_x$ and width $L_y$, with the center at origo, and in contact with (infinitely long) homogeneous normal conductors of width $L_y$ at the two boundaries where $x=\pm L_x/2$.

One approximation to the problem corresponds to using the standard GL equations and a linear function for the magnetic field at the boundary \cite{MachidaPRL1993}
\begin{equation}
\mathbf{B}^{(e)}=\mathbf{B}^{(a)}+  {B}^{(c)} \frac{2}{L_y} y \: \mathbf{e}_z. \label{linH}
\end{equation}
This corresponds to using the same current-induced magnetic field as that from a superconductor with a constant current density.
This assumption does not take into account the interplay between the dynamics of the currents and the magnetic field at the boundary of the superconductor.
The true currents are larger close to boundaries and vary substantially due to the local response of applied fields and correlates with the vortex dynamics present.
In order to focus on the role of vortex dynamics, we set $\mathbf{B}^{(a)}=0$ in the following.
Let us show here how  Eq.~(\ref{linH}) is then related to the standard BC of  Eqs.~(\ref{VSBC})  and~(\ref{MSBC}) and that they are redundant. 
The third BC in Eq.~(\ref{MSBC}) can be reformulated to $\mathbf{A} = - \mathbf{J}^{(e)}  / \sigma \cdot t $ where the current goes through the boundaries.
However, this BC needs in fact not to be explicitly implemented, since the following relation between the second and third BC in Eq.~(\ref{MSBC}) holds for the left and right boundary
\begin{equation}
\mathbf{A} = -  \frac{ I}{ \sigma L_y } t \: \mathbf{e}_x \  \Leftrightarrow \   \nabla \times \mathbf{A}   =  \frac{I}{L_y}y \: \mathbf{e}_z .\label{eq:THIRD_B.C.}
\end{equation}
Indeed, the right form of the above BC gives a current density at the left- and right-boundaries that are parallell to  $\mathbf{e}_x$, since according to \emph{Ampere's law} for the current density (assuming a static electric displacement field $\partial_{t}\mathbf{D}=0$)
\begin{equation}
\mathbf{J} \left( x=\pm  L_x/2,y \right) =\nabla\times \mathbf{B} =\frac{I}{L_y} \: \mathbf{e}_{x} ,
\end{equation}
where
\begin{equation}
I = \int^{L_y/2}_{-L_y/2}  \mathbf{J}^{(e)}  \cdot \mathbf{e}_x \: dy ,  \   \mathbf{J}^{(e)}  =\frac{I}{L_y} \:  \mathbf{e}_x. \label{Je}
\end{equation}
Hence, one can implicitly reproduce the third BC of Eq. (\ref{MSBC}) by a proper choice of the magnetic field at the boundaries.
Whatever choice is made from Eq.~(\ref{eq:THIRD_B.C.}) for the left and right BC of the rectangle, the upper and lower BC should now fulfill
\begin{equation}
\mathbf{B}^{(e)} \left( x, y=\pm L_y/2 \right) = \mathbf{B}^{(c)} = \pm I/2 \label{Idiv2}.
\end{equation}
We have confirmed in numerical calculations that we can equivalently use either of the two possibilities in  Eq.~(\ref{eq:THIRD_B.C.}) for the left and right BC, and Eq.~(\ref{Idiv2}) for the upper and lower BC. 

In order to treat the BC self-consistently, the magnetic field at the boundaries should be calculated from the true current density and coupled into the GL equations as formulated in Eqs.~(\ref{GLE}).
In the following we discuss this general problem.
We use  the left part of Eq.~(\ref{eq:THIRD_B.C.}), i.e. we assume that the non-trivial current dynamics inside the superconductor do not affect the normal conductors that it is connected to,
and $\psi = 0$ for the left and right BC.
Then we solve the lower equation of~(\ref{GLE}), with $ \mathbf{A} \cdot\mathbf{n}=0$ and $\nabla\psi\cdot\mathbf{n}=0$ for the upper and lower boundaries. 
The dynamics of EM-fields in the surrounding half-spaces are not considered.

\section{Numerical results}
\label{}

\begin{figure}[tbp]
\vspace{3mm}
\hspace{6mm}  {\bf (a)}
\begin{center} 
\vspace{-8mm}
\includegraphics[height=41mm]{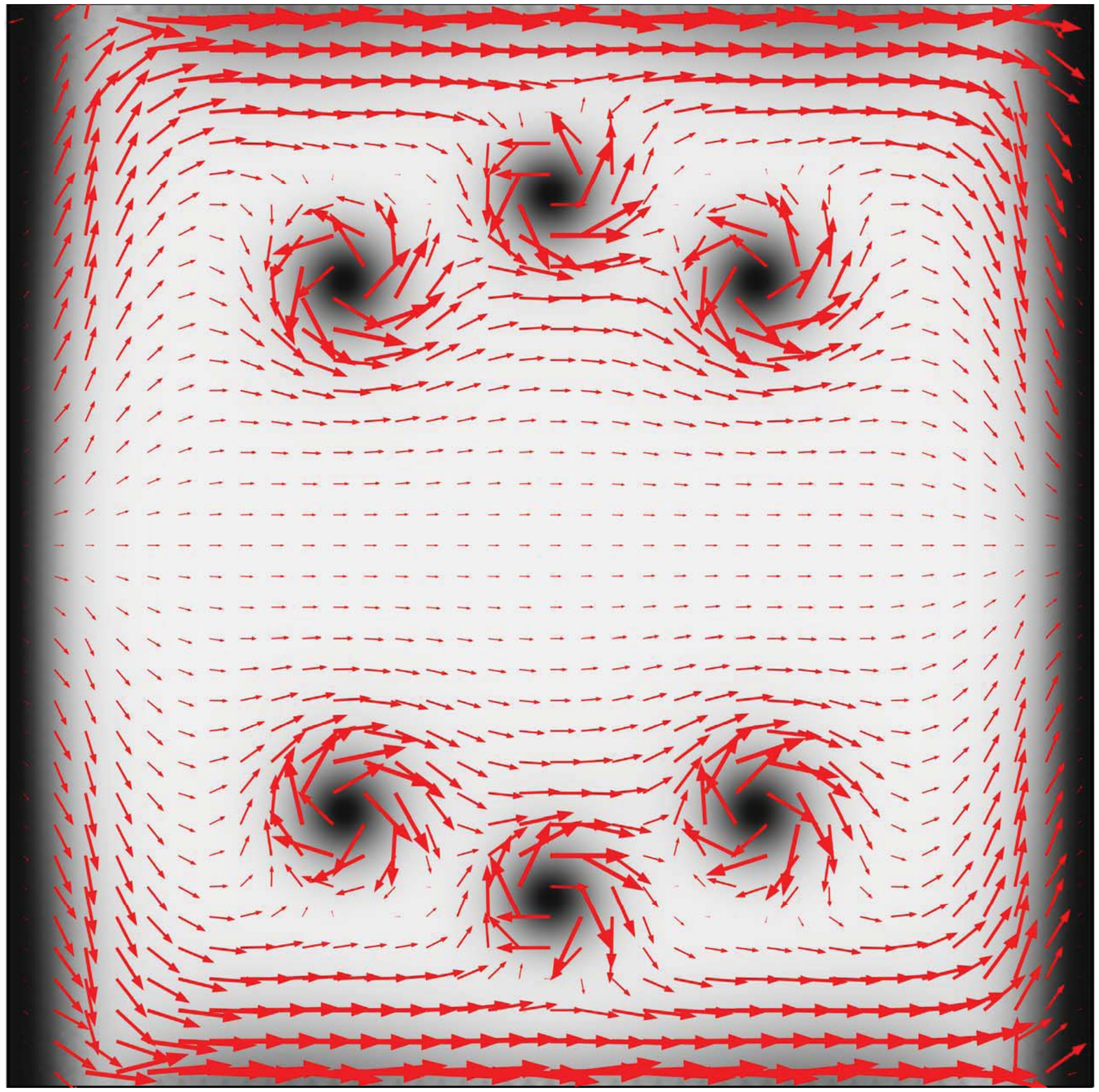} 
\end{center}
\vspace{-1mm}
\vspace{3mm}
\hspace{6mm}  {\bf (b)}
\begin{center} 
\vspace{-8mm}
\includegraphics[height=44mm]{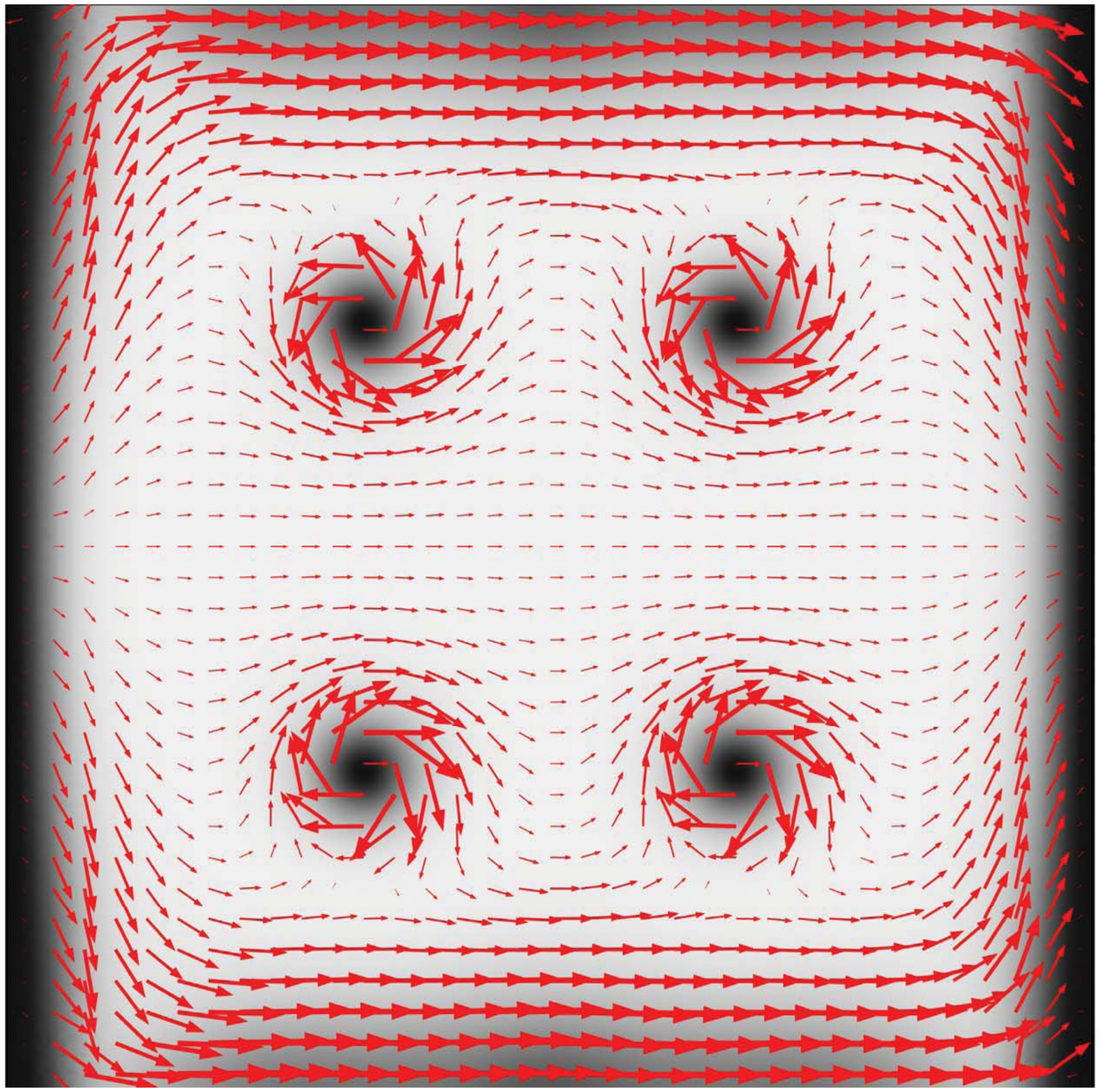} 
\end{center}
\vspace{-1.5mm}
\caption{ Super-currents $ \mathbf{J}^{(s)} $ (arrows) of Eq.~(\ref{Js}) for; (a) a self-consistent-; and (b) a linear-; calculation of the current-induced magnetic field.
Background gray-scale shows the corresponding density $ \left|\psi\right|^{2}$ of Cooper-pairs.
Parameters used are $\sigma= \kappa=4$ (type-II), $I=1.55$ and $ \mathbf{B}^{(a)}=  0$.
The size of the rectangle is $L_x=L_y=10 \lambda$.
The dimensionless time shown here is $t=200$, however, the external current $I$ and the amplitude of the self-consistent BC are applied gradually to keep it numerically tractable. 
At  $t=200$ it is $8 \%$ of the self-consistent amplitude that is mixed into the BC, which is already sufficient to demonstrate qualitative differences.
Convergence has been confirmed with different meshes.
}
\label{figure1}
\end{figure}

We use a FEM implementation with cubic Lagrange shape functions and a time-dependent adaptive mesh refinement \cite{Comsol} to solve Eqs. (\ref{GLE}).
The total transport current through the superconductor is monitored numerically with a projection onto  $\mathbf{e}_x$ (direction of transport) with integration along lines orthogonal to $\mathbf{e}_x$ (i.e. $I=\int J_x (x,y) dy$). 

We give examples of snapshots of the super-currents and vortex structure for a self-consistent implementation of Eqs. (\ref{GLE}) in Fig.~\ref{figure1}~(a),
and for a simulation, using the approximate BC of  Eq.~(\ref{linH}), in Fig.~\ref{figure1}~(b).
On the boundary $ \mathbf{r} \in \partial \Omega $ we use a simplified notation for the magnetic field at the upper-/lower- ($U/L$) boundaries of the rectangle
\begin{equation}
B_{U/L}  =  B_{U/L}^{(SC)}  + B_{U/L}^{(NC)}  \equiv \mathbf{B}^{(c)} \left( \! x, y=\pm   \frac{L_y}{2} \right) \! \cdot \! \mathbf{e}_{z} .   \label{defB_UL}
\end{equation}
In practice we implement a numerical integration over the superconductor for each time-step.
For the trivial rectangular geometry, for example 
\begin{equation}
 B_U^{(SC)} = \frac{ 1}{4 \pi } \! \int \! \! \! \! \int_{\Omega} \frac{ \! \! J_x  \left(\frac{L_y}{2} \! \! - \! \! y' \right) \! - \!  J_y  \left(x \! \! - \! \! x' \right) }{  \left( \left( x \! \! - \! \! x'\right)^2 +  \left(\frac{L_y}{2} \! \! - \! \! y'\right)^2 \right) ^{ \! \! 3/2}   } dx' dy', \label{h1}
\end{equation} 
is the contribution to $B_U$ from the superconductor.
Then $B_L^{(SC)}$ is obtained by an analogue construction.
The contribution $B_U^{(NC)}$ from the integration over the two semi-infinite normal conductors,
where $ \mathbf{J}^{(e)} = I/L_y \: \mathbf{e}_x$, is an even function in $x$, odd in $y$, such that $B_L^{(NC)} = -  B_U^{(NC)}$, and needs to be calculated only once
\begin{eqnarray}
\label{h1tilde}
\begin{array}{l}
\frac{4 \pi L_y}{I} B_U^{(NC)} =  \ln \left(  \frac{L_x^2}{L_x^2 -4x^2} \right)  
\\
+ \textnormal{atanh}\! \! \left( \! \!  \frac{ \frac{L_x}{2}+x}{ \sqrt{ L_x^2 +  \left(\frac{L_x}{2}+x \right)^2 }  } \! \! \right) \! \! + \! \! \textnormal{atanh} \! \!  \left( \! \! \frac{\frac{L_x}{2}-x}{\sqrt{L_x^2 +  \left(\frac{L_x}{2}-x \right)^2 } } \! \! \right)    .
\end{array}
\end{eqnarray}
\begin{figure}[tbp]
\begin{center} 
\includegraphics[height=32mm]{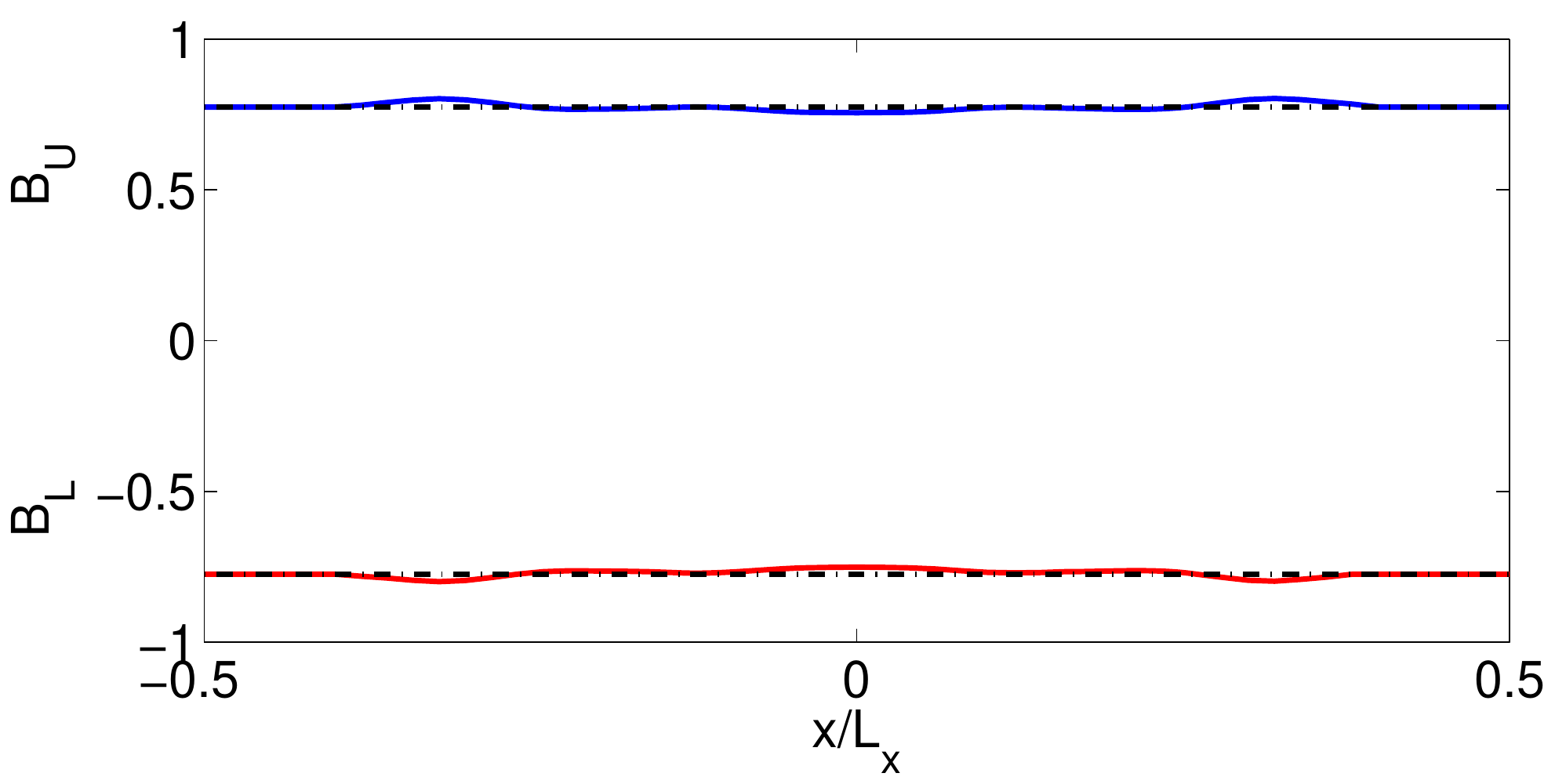} 
\end{center}
\vspace{-3mm}
\caption{Current-induced magnetic fields at the upper and lower boundaries corresponding to  Fig.~\ref{figure1} (a), solid curves shows results of Eqs.~(\ref{defB_UL})-(\ref{h1tilde}).
Dashed lines shows the approximate theory of Eq.~(\ref{Idiv2}).
Parameters as in Fig.~\ref{figure1}.}
\label{figure2}
\end{figure}
In Fig.~\ref{figure2} $B_{U/L}$ from Eqs.~(\ref{defB_UL})-(\ref{h1tilde}) is compared with the linear approximate theory of Eq.~(\ref{linH}).
In the numerical time evolution
we observe the creation of antivortex-vortex pairs symmetrically at the upper and lower boundaries (Fig.~\ref{figure1}), followed by annihilation within the superconductor when they meet at the horizontal axis. 
Note in particular that $B_L = - B_U$ here (Fig.~\ref{figure2}) due to the symmetry of antivortex-vortex pairs in the absence of an external applied magnetic field.
If there is also an applied field  $\mathbf{B}^{(a)} \neq 0$,
vortices are created at the boundary with highest external magnetic field  (if above a critical value). 
 Generally  $B_{U/L}$ then doesn't obey any symmetries but reflects the non-trivial dynamics of external fields and vortices present.

Following the positions of vortices and current-induced fields through time, one can qualitativelly explain observed differences between the simulations using self-consistent vs linear BC [i.e. Figs.~\ref{figure1}~(a)~vs~(b)].
We here refer to those as simulations (a) and (b), respectively.
For example, the three vs two antivortex-vortex pairs present at $t=200$ in Figs.~\ref{figure1}~(a) and~(b), can be explained as a memory effect for $t \sim 145$ (not shown) when vortices communicates with the BC via Eq.~(\ref{h1}).
At this point in time, both simulations are similar, due to the gradual increase of the amplitude for the self-consistent BC, and show three antivortex-vortex pairs.
For the (b) simulation, two more pairs then enter such that a short time before the three central pairs annihilate, there are five pairs present simultaneously.
However, for simulation (a), the three pairs present at  $t \sim 145$  in effect reduce the current-induced fields, since a vortex introduces an opposite directed (to $ \mathbf{e}_x$) super-current with a smaller distance to the (e.g.) $y=L_y/2$ boundary compared to the lower part of the same vortex, and hence locally lower the value of Eq.~(\ref{h1}).
New antivortex-vortex pairs in simulation (a) are therefore delayed in time until older pairs have been annihilated.

\section{Summary}
\label{}
We have presented self-consistent boundary-conditions to model transport currents with the time-dependent Ginzburg-Landau equation.
Numerical results for the vortex dynamics confirm qualitative differences compared to the standard GL equation and BC, however, a practical drawback is a substantially higher computational cost.

\section*{Acknowledgements}
\label{}
We thank V. Rodriguez-Zermeno and D. Sj\"{o}berg for useful discussions. M.~\"{O}. is supported by H.~C. \O rsted and ESF grants.





\bibliographystyle{elsarticle-num}
\bibliography{<your-bib-database>}

\begin{thebibliography}{00}
\bibitem{Comsol} The software Comsol 4.2 have been used for most numerical calculations. 
\end{thebibliography}



\end{document}